\documentclass[preprint, 3p]{elsarticle}

\usepackage{amssymb}

\usepackage{siunitx}
\usepackage{graphicx}
\usepackage{caption}
\usepackage{subcaption}
\usepackage[table]{xcolor}
\usepackage{booktabs}
\usepackage[colorlinks=true]{hyperref}
\usepackage{bm}

\journal{Journal}

\begin{document}

\begin{frontmatter}



\title{Investigating the Applicability of a Snapshot Computed Tomography Imaging Spectrometer for the Prediction of \textdegree Brix and pH of Grapes}

\author[add1,add2]{Mads Svanborg Peters}
\author[add3,add6]{Mads Juul Ahlebæk}
\author[add3,add6]{Mads Toudal Frandsen}
\author[add2]{Bjarke Jørgensen}
\author[add3]{Christian Hald Jessen}
\author[add4]{Andreas Krogh Carlsen}
\author[add2]{Wei-Chih Huang}
\author[add1,add6]{René Lynge Eriksen}

\address[add1]{SDU NanoSyd, Mads Clausen Institute, University of Southern Denmark}
\address[add2]{Newtec Engineering A/S, Denmark}
\address[add3]{Department of Physics, Chemistry and Pharmacy, University of Southern Denmark}
\address[add6]{SDU Climate Cluster, the University of Southern Denmark}
\address[add4]{Mads Clausen Institute, University of Southern Denmark}

\begin{abstract}
In this paper, a recently developed snapshot hyperspectral imaging (HSI) system based on Computed Tomography Imaging Spectroscopy (CTIS) is utilized to determine \textdegree Brix and pH values in \textit{Sheegene 20} table grapes through Partial Least Squares Regression (PLSR) modeling. The performance of the CTIS system is compared with that of a state-of-the-art line scan HSI system by imaging 100 grapes across both platforms. Reference measurements of \textdegree Brix and pH values are obtained directly using a refractometer and a pH meter, as these parameters are essential for assessing the quality of table and wine grapes. The findings indicate that the spectra captured by the CTIS camera correlate well with the reference measurements, despite the system's narrower spectral range. The CTIS camera’s advantages, including its lower cost, portability, and reduced susceptibility to motion errors, highlight its potential for promising in-field applications in grape quality assessment.

\end{abstract}




\end{frontmatter}

\section{Introduction}\label{sec:introduction}
Pre-harvest grape maturity and quality assessment are of utmost importance in the table grape and wine grape industry, as they directly influence the taste and aroma of the final product. Measurable indicators of maturity and quality in the lab include pH, titratable acidity (TA), total phenols and total anthocyanins, and soluble solids content (SSC) measured in \textdegree Brix, which represents the sugar content in the grape juice.
For example, the \textdegree Brix/TA ratio was found to be a good indicator of the optimum harvesting stage of a number of different table grape varieties in~\cite{jayasena_brixacid_2008}. 

Traditionally, \textdegree Brix values are measured using a refractometer, which requires extracting a sample of the grape juice. This process is time-consuming, and labor-intensive, it can introduce errors due to variations in juice extraction and may not capture all inter-row or inter-vine variations. 
Therefore it is of great interest to replace or complement these measurement methods with non-invasive spectroscopy~\cite{cozzolino_analysis_2006} and hyperspectral imaging (HSI)~\cite{baiano_application_2012}. 

Compared with traditional spectroscopy, HSI enables a two dimensional (2D) analysis of the reflectance spectra across the entire grape skin area, providing more information about the grape's chemical composition including its SSC and pH. An early study of pH, \textdegree Brix and TA in table grapes, of the variety Crimson Seedless, based on line scan hyperspectral images (hereafter referred to as datacube) was presented in
\cite{baiano_application_2012}. Several studies have since demonstrated the potential of line scan HSI for predicting \textdegree Brix values in table- and wine grapes~\cite{ram_systematic_2024}. 

An extensive study of different predictive data models that can be used to extract predictions for oenological parameters based on HSI is presented in~\cite{gomes_determination_2021}. 

All of the above studies employ line scan hyperspectral cameras with grapes put on a conveyor belt. The purpose of this paper is to compare the performance of a snapshot hyperspectral Computed Tomography Imaging Spectrometer (CTIS) with a line scan hyperspectral camera. The CTIS captures spatial and spectral information simultaneously by utilizing a 2D diffraction grating to disperse the light into a $3\times3$ diffraction pattern of a central zeroth order and 8 surrounding first orders. From this diffraction image, a tomographic reconstruction is required to obtain a hyperspectral datacube \cite{descour_computed-tomography_1995,Th,Okamoto:91}. The motivation for this comparison includes the ease of in-field deployment of the CTIS camera and the snapshot ability which means it will be less prone to motion errors when later deployed in the field~\cite{ram_systematic_2024}.

We employ a newly developed CTIS system with a spectral range from 600-850 nm. We capture snapshot datacubes of 100 different grapes of the variety \textit{Sheegene 20}, reconstruct the grape datacubes using both the standard expectation maximization (EM) algorithm and a U-Net neural network, and develop predictive Partial Least Squares Regression (PLSR) models that correlate the spectral information with corresponding \textdegree Brix- and pH measurements, acquired using a refractometer and a pH-meter, respectively. 

We compare the CTIS system performance with a state-of-the-art line scan HSI system with a spectral range from 532-1655 nm. The results of this study have the potential to improve the efficiency and accuracy of grape quality assessment in the wine industry and could have broader applications in the agriculture and food industries.

\section{Methods}\label{sec:methods}
\begin{figure}[h]
    \centering
    \includegraphics[width = \textwidth]{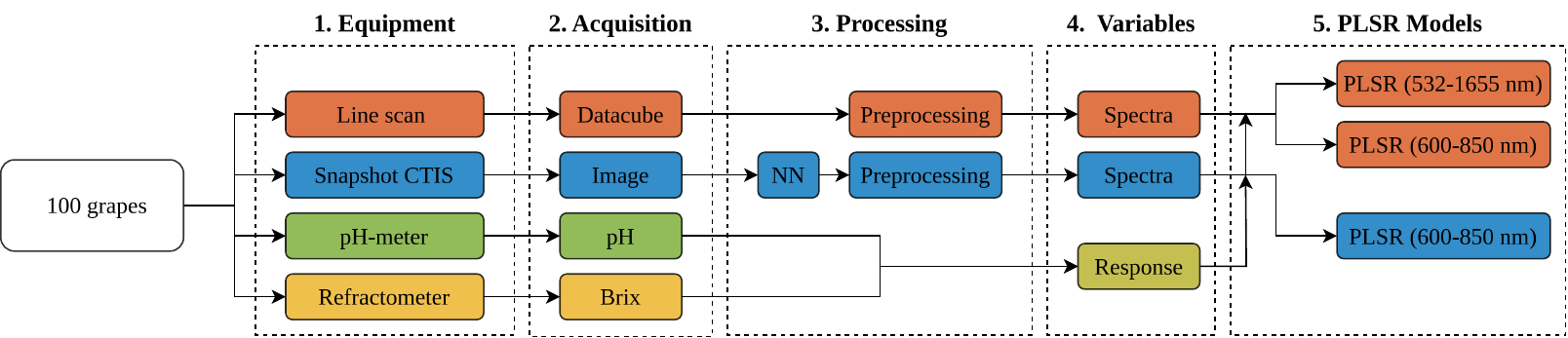}
    \caption{Flowchart of the data pipeline from data acquisition to PLSR models for the snapshot CTIS and line scan system. For each grape, a line scan datacube, snapshot CTIS image, pH measurement and \textdegree Brix measurement are acquired. The datacubes are preprocessed and the mean standardized spectrum (expressed by SNV) of each grape is determined. PLSR models are created from the mean spectra and the measured response variables (pH value and \textdegree Brix number). }
    \label{fig:Overview}
\end{figure}
Fig.~\ref{fig:Overview} provides an overview of the data acquisition and analysis pipelines for equipment and measured variables used to create the PLSR models. The overview is divided into five sections: Equipment, Data Acquisition, Processing, Variables, and PLSR Models. In total 100 \textit{Sheegene 20} table grapes were used, purchased at commercial harvest ripeness from a local fruit market. 

For the line scan HSI system (orange), the captured data is preprocessed to obtain mean grape spectra, serving as predictor variables. These are combined with response variables from the pH meter (green) and refractometer (yellow).

In the case of the CTIS system (blue), a neural network (NN) reconstructs the 2D captured image into a 3D datacube $\bm{I}(x,y,\lambda)$, where the intensity $\bm{I}$ is a function of spatial coordinates $(x,y)$ and wavelengths $\lambda$, which is further preprocessed to obtain mean grape spectra averaged over the visible grape surface. Similar to the line scan pipeline, these spectra are used as predictor variables, along with the same pH-meter (green) and refractometer (yellow) response variables as in the line scan pipeline.





\subsection{Experimental setup for hyperspectral imaging}\label{subsec:setup}
\begin{figure}
    \centering
    \begin{subfigure}[t]{0.49\textwidth}
        \centering
        \includegraphics[width=\textwidth]{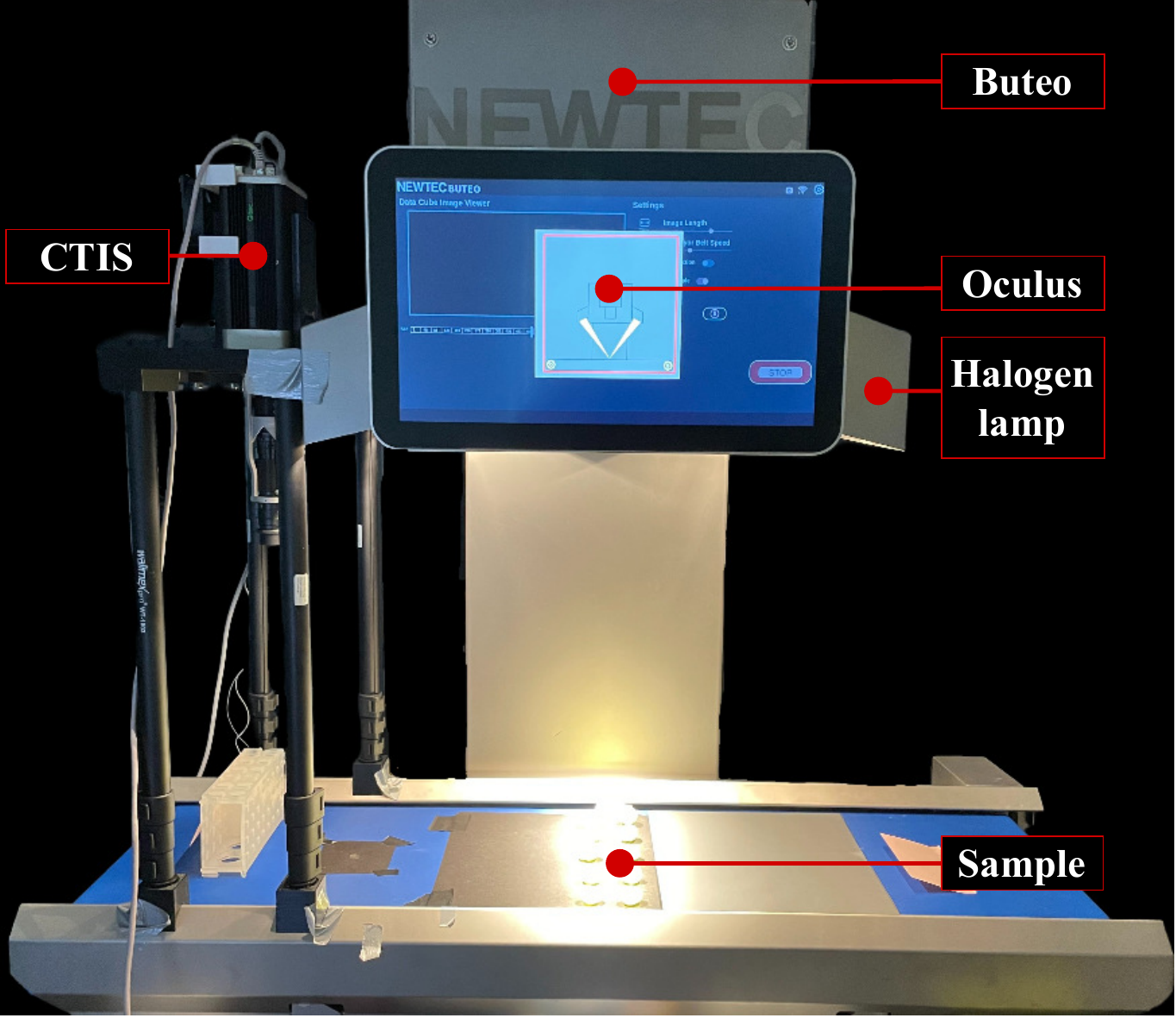}
        \caption{Experimental setup}
        \label{fig:experimental_setup}
    \end{subfigure}
    \hfill
    \begin{subfigure}[t]{0.49\textwidth}
        \centering
        \includegraphics[width=\textwidth]{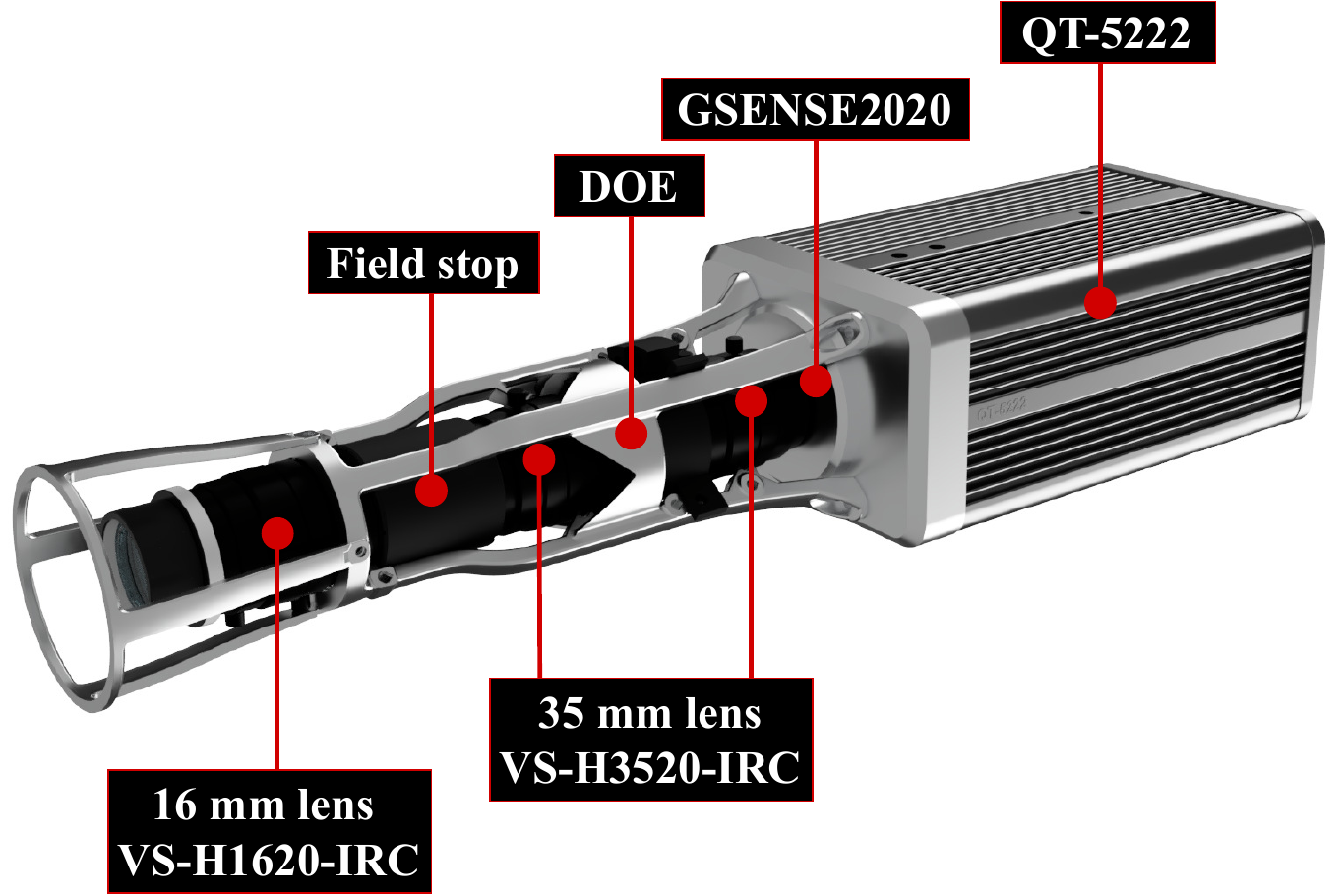}
        \caption{CTIS}
        \label{fig:CTIS_system}
    \end{subfigure}
    \caption{Experimental setup for hyperspectral imaging of grapes with both line scan (Buteo) and snapshot (CTIS) HSI systems. }
    \label{fig:setup_new}
\end{figure}

The experimental setup \textcolor{blue}{(Fig.~\ref{fig:experimental_setup})} for the HSI consists of a Newtec Buteo line scan HSI system~\cite{noauthor_pushbroom_2023} and a newly developed snapshot hyperspectral CTIS camera \textcolor{blue}{(Fig.~\ref{fig:CTIS_system})}.
The Buteo consists of a conveyor belt, 14 halogen lamps positioned above the conveyor at ± 45\textdegree \ angles, and a Newtec Oculus Vis-SWIR hyperspectral line scan camera \textcolor{blue}{(Fig.~\ref{fig:experimental_setup})}. The Oculus camera houses a 1.34 MP IMX990 Sony SWIR image sensor and uses a 12 mm ViSWIR Hyper APO lens from Computar. The line scan system captures 900 spectral channels in the spectral range from 445-1710 nm at a spatial resolution of $1296 \times m$ pixels - where $m$ is the scan length in pixels - corresponding to $0.25 \text{ mm/pixels} \times m $ at a working distance at 600 mm.

A 3D-printed PETG mount is used to fix the CTIS camera above the conveyor belt of the Buteo. The CTIS camera is an upgraded version of the system described in~\cite{peters_high-resolution_2022}. It houses a monochrome 4 MP GSENSE2020 CMOS sensor, a custom diffractive optical element (DOE), two 35 mm Vis-NIR VS-technology lenses (VS-H3520-IRC), and a 16 mm Vis-NIR VS-technology lens (VS-H1620-IRC) \textcolor{blue}{(Fig.~\ref{fig:CTIS_system})}. The spectral range of the camera from 600-850 nm is defined by a 600 nm longpass (FELH0600) and an 850 nm shortpass (FESH0850) filter from Thorlabs at the front of the system. The system outputs hyperspectral datacubes with spatial dimensions of $312\times \SI{312}{pixels}$ and 236 or 45 spectral channels depending on whether the reconstruction is performed by the EM algorithm or the U-Net~\cite{navab_u-net_2015}, respectively. The U-Net is based on our previous work with Convolutional Neural Network (CNN)-based reconstruction of CTIS images~\cite{huang_application_2022,ahlebaek_hybrid_2023}, and outputs hyperspectral datacubes as relative reflectance values. The reconstruction of CTIS images is detailed in Sec.~\ref{subsec:acq_nn}


\subsection{Acquisition of hyperspectral images}\label{subsec:acq_Buteo}
The line scan acquisition of hyperspectral images was performed with ten grapes at a time: The ten grapes were placed on matte black cardboard with similar reflectance to prevent overexposure of either the background or the grapes and ensure high contrast between them. The grapes were held down with double-sided tape to minimize movement during the scanning. Due to the low reflectance of the grapes, a gray reference target is placed next to the grapes to remove illumination variations (flatfield correction) and to transform the captured intensities to reflectance. The gray reference target is calibrated against a RESTAN white reference target~\cite{image_engineering_gmbh__co_kg_restan_2024}. The camera exposure time and gain were set to maximize the dynamic range of the system with minimal noise. Dark frames were captured for the chosen camera settings. This process was performed as presented in~\cite{gomes_determination_2021}. The grapes were imaged in three different orientations  (0\textdegree, 120\textdegree \ and 240\textdegree) by rotating the grapes 120\textdegree \ around the major axis between captures. The additional orientations are utilized to be less prone to marks or defects on a single side of a grape. Thus, 30 line scan datacubes are captured, each containing 10 grapes corresponding to 100 grapes imaged in three different rotations.

Immediately after capturing the line scan datacubes (all three orientations), the grapes on the cardboard paper were individually imaged by the CTIS camera under illumination from the halogen lamps of the line scan system. The camera exposure time and gain were set to maximize the dynamic range of the system, and dark frames were captured and subtracted from the captured images. Similarly, the grapes were rotated 120\textdegree \ between captures to obtain images for the three orientations. Thus, 300 CTIS images were captured corresponding to 100 grapes imaged in three different orientations. 

\subsection{Measurement of grape \textdegree Brix and pH}\label{subsec:acq_meas}
Directly after image acquisition, each grape was subjected to the determination of acidity and soluble solid content (SSC) expressed in pH and \textdegree Brix, respectively. Each grape was squashed individually and the juice was collected for the acidity measurement using a Metrohm 913 pH-meter (with a documented accuracy of $\pm \ 0.003$ pH), while the SCC was measured by a Milwaukee MA885 Digitial Wine Refractometer (with a documented accuracy of $\pm \ 0.2$ \textdegree Brix. Between each measurement, all equipment was thoroughly cleaned using demineralized water and wiped using laboratory tissues.

\subsection{Preprocessing of hyperspectral images}
\begin{figure}
    \centering
    \hfill
    \begin{subfigure}[t]{0.1844\textwidth}
        \centering
        \includegraphics[width=\textwidth]{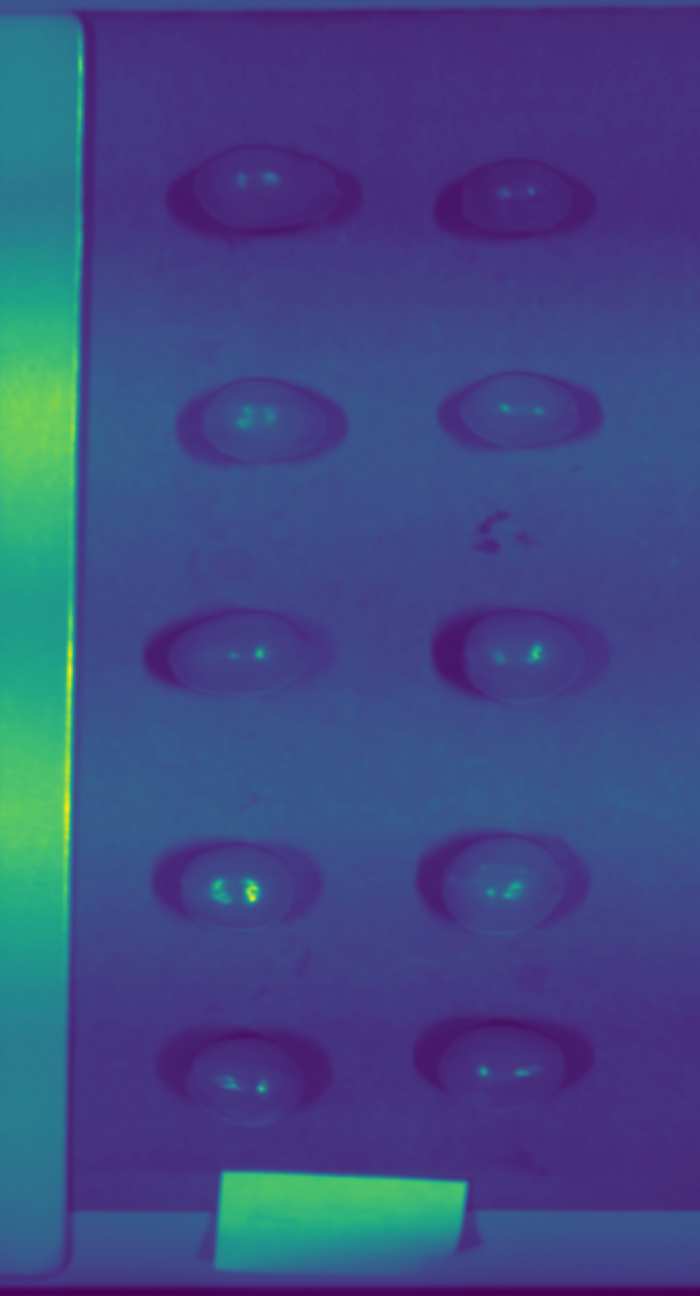}
        \caption{Raw}
        \label{fig:buteo_preproc_a}
    \end{subfigure}
    \hfill
    \begin{subfigure}[t]{0.1807\textwidth}
        \centering
        \includegraphics[width=\textwidth]{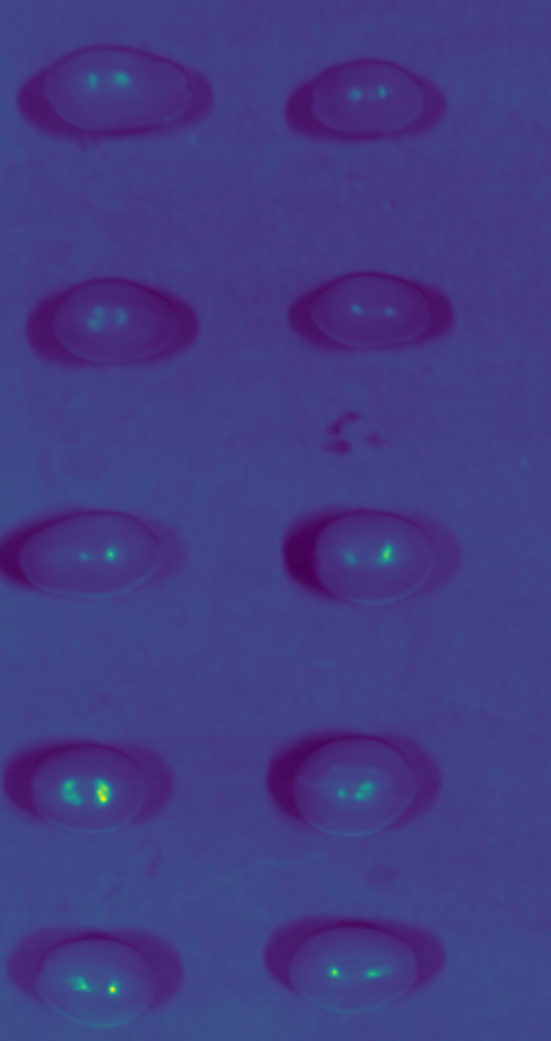}
        \caption{Reflectance}
        \label{fig:buteo_preproc_b}
    \end{subfigure}
    \hfill
    \begin{subfigure}[t]{0.1807\textwidth}
        \centering
        \includegraphics[width=\textwidth]{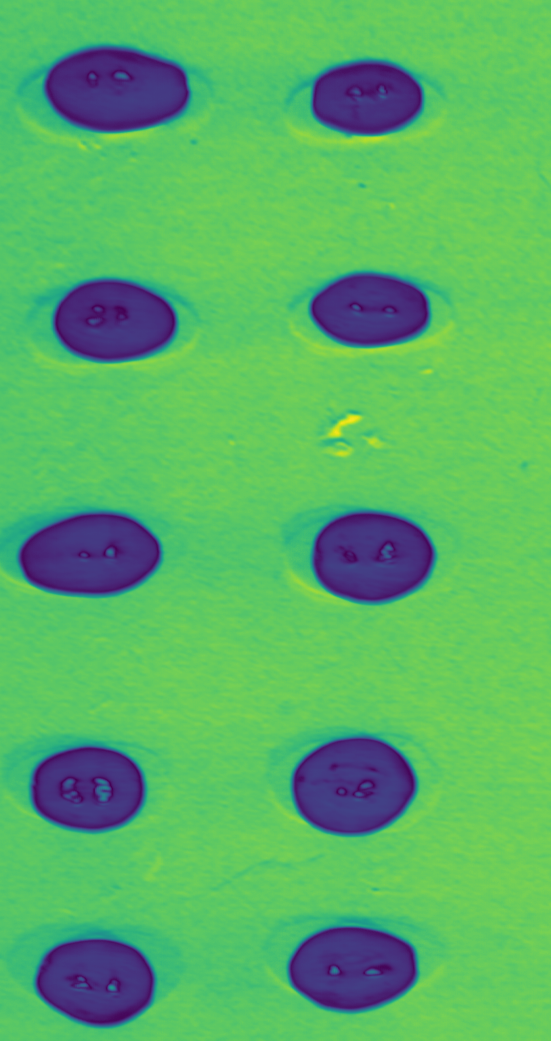}
        \caption{SAM score}
        \label{fig:buteo_preproc_c}
    \end{subfigure}
    \hfill
    \begin{subfigure}[t]{0.2013\textwidth}
        \centering
        \includegraphics[width=\textwidth]{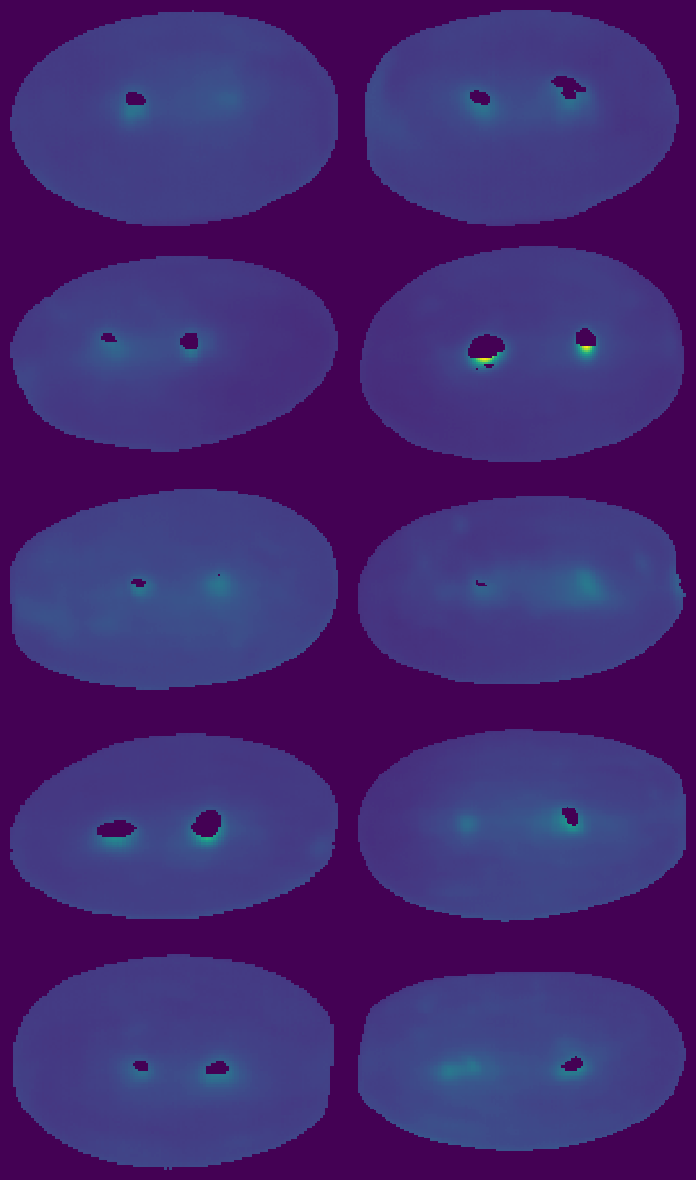}
        \caption{Isolated grapes}
        \label{fig:buteo_preproc_d}
    \end{subfigure}
    \hfill
    \begin{subfigure}[t]{0.2013\textwidth}
        \centering
        \includegraphics[width=\textwidth]{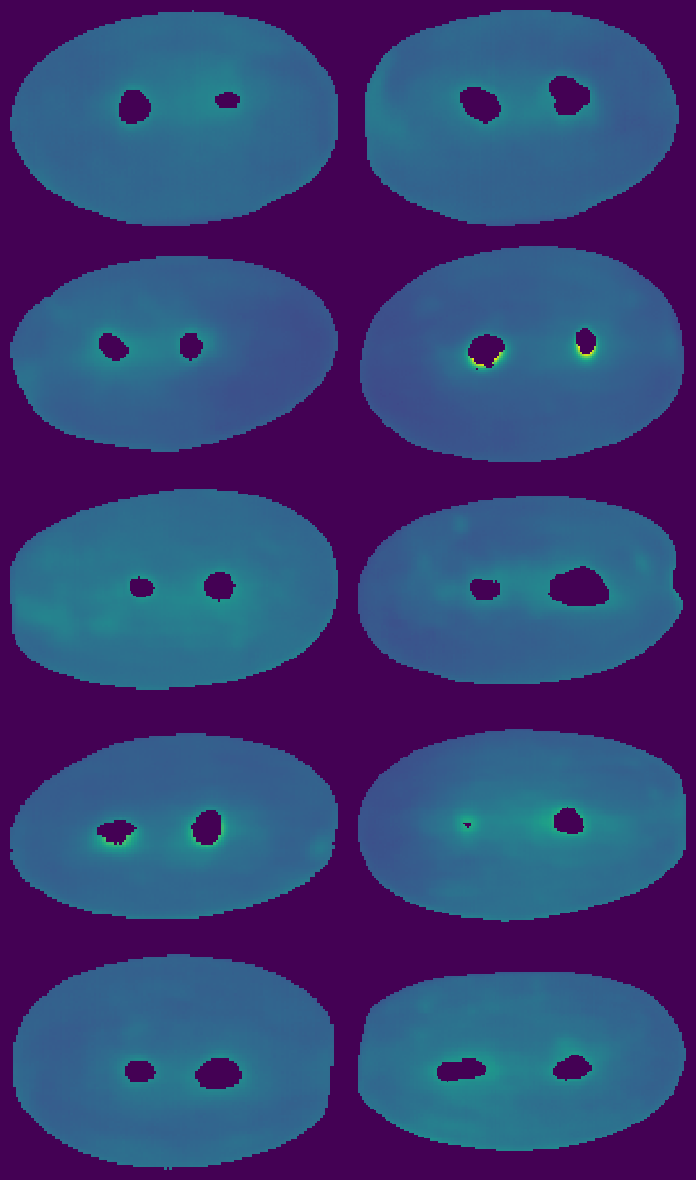}
        \caption{Reflections removed}
        \label{fig:buteo_preproc_e}
    \end{subfigure}
    \hfill
    \caption{Data preprocessing pipeline for line scan hyperspectral images: (a) Mean raw datacube, (b) reflectance datacube cropped to region with grapes, (c) SAM score map for datacube and mean reflectance spectrum of a grape, (d) isolated grape datacubes by SAM score and thresholding, and (e) final grape datacubes after removal of reflections. All images are illustrated using the viridis colormap.}
    \label{fig:buteo_preproc}
\end{figure}

The data preprocessing pipelines differ between the line scan and snapshot systems. For the line scan datacubes, the raw captured cube (Fig.~\ref{fig:buteo_preproc_a}) is first transformed to reflectance by subtraction of the dark frame $\bm{D}$ and elementwise division ($\oslash$) by the column-mean white reference target intensity $\bm{W}$:
\begin{equation}\label{eq:reflectance}
    \bm{R}(x,y,\lambda) = \left(\bm{I}(x,y,\lambda)-\bm{D}(x,y,\lambda)\right) \oslash \frac{1}{N_y} \sum_{y=0}^{N_y} \left( \bm{W}(x,y,\lambda)-\bm{D}(x,y,\lambda) \right)
\end{equation}
where $N_y$ is the number of columns that the column-mean is calculated over. Subsequently, the reflectance cube is cropped to remove the gray reference on the left (Fig.~\ref{fig:buteo_preproc_b}).

The ten grapes are isolated using the mean spectrum of a grape (manually determined) and the spectral angle mapper (SAM) algorithm is used. The SAM algorithm determines the spectral similarity between a reference spectrum and a test spectrum by calculating the $\lambda$-dimensional angle between the two spectra. The SAM algorithm is defined as:

\begin{equation} \label{eq:SAM}
    \alpha=\cos^{-1}\left(\frac{\sum_{i=1}^{\lambda} t_i r_i}{\sum_{i=1}^{\lambda} t_i^2 \sum_{i=1}^{\lambda} r_i^2}\right)
\end{equation}
Where $\alpha$ is the resulting angle between the test $t_i$ and the reference $r_i$, while $\lambda$ is the number of wavelengths.

A low $\alpha$ means high spectral similarity. Using SAM for all spectra in the datacube, results in a distance map (Fig.~\ref{fig:buteo_preproc_c}) from which a mask for the grapes is computed through thresholding. 
Subsequently, the datacube is partitioned into ten smaller grape-containing datacubes (Fig.~\ref{fig:buteo_preproc_d}), and the spectrum for each grape is averaged. To remove grape reflections, we create an additional SAM distance map using the mean spectrum of the grapes with reflections, update the masks, and with the updated masks recompute the mean spectrum of the grapes without reflections (Fig.~\ref{fig:buteo_preproc_e}). This process is repeated for all three orientations of each grape. In the end, a single mean spectrum for each grape is obtained by averaging across the three rotation-specific mean spectra.


\begin{figure}
    \centering
    \hfill
    \begin{subfigure}[t]{0.32\textwidth}
        \centering
        \includegraphics[width=1\textwidth]{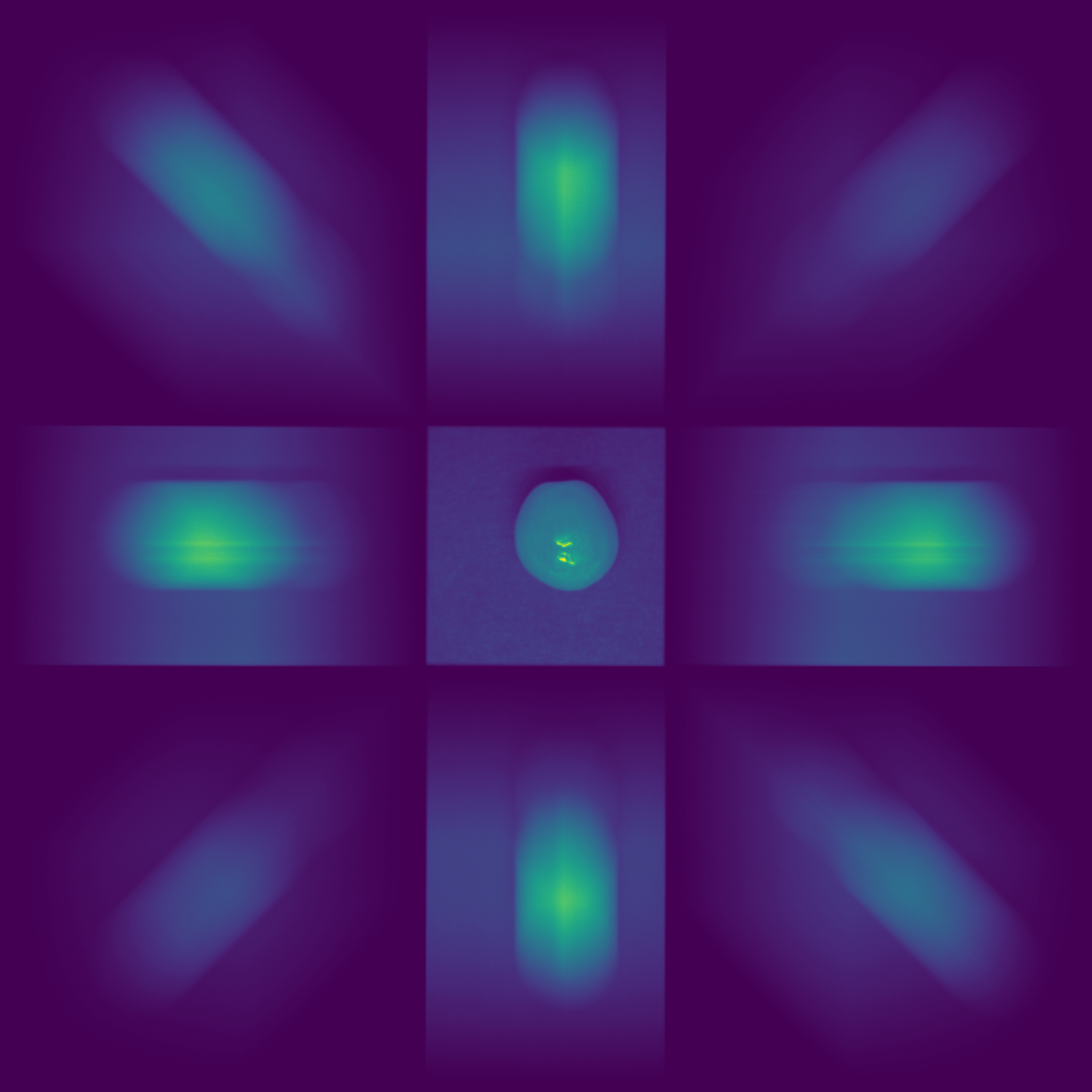}
        \caption{CTIS image}
        \label{fig:CTIS_img}
    \end{subfigure}
    \hfill
    \begin{subfigure}[t]{0.32\textwidth}
        \centering
        \includegraphics[width=1\textwidth]{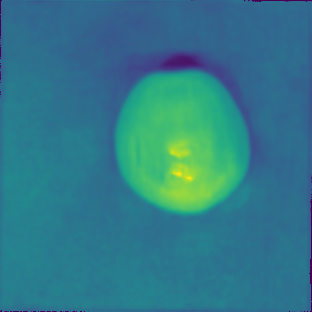}
        \caption{Mean reconstructed datacube}
        \label{fig:CTIS_recon}
    \end{subfigure}
    \hfill
    \begin{subfigure}[t]{0.32\textwidth}
        \centering
        \includegraphics[width=1\textwidth]{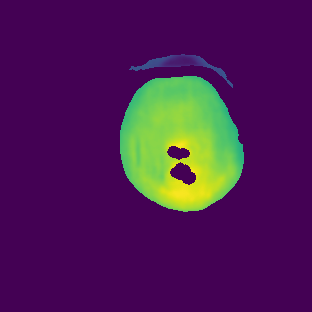}
        \caption{Masked datacube}
        \label{fig:CTIS_Erod}
    \end{subfigure}
    \hfill
    \caption{Data preprocessing pipeline for snapshot hyperspectral images. (a) Cropped and dark frame subtracted CTIS image, (b) mean reconstructed datacube (along spectral dimension), and (c) masked reconstructed datacube from which the mean spectrum is calculated. All images are illustrated using the viridis colormap.}
    \label{fig:CTIS_pipeline}
\end{figure}

For the snapshot images; initially, five dark-frame CTIS images were captured and averaged, and subsequently used to dark-frame correct all acquired CTIS images (Fig.~\ref{fig:CTIS_img}). This is done to account for possible dead pixels and to eliminate sensor bias, effectively leading to more accurate datacube reconstructions. The dark frame-corrected CTIS images are then fed into a neural network (Section \ref{subsec:acq_nn}) that outputs reconstructed datacubes (Fig.~\ref{fig:CTIS_recon}).
The grapes are isolated by calculating an Euclidian distance map between the mean background spectrum and the remaining datacube. From the distance map a mask for the grapes is computed through thresholding. Reflections and edge effects are removed by eroding the binary mask using a spherical structure element (Fig.~\ref{fig:CTIS_Erod}). 
As for the line scan datacubes, the rotation-specific mean spectra are averaged into a single mean spectrum for each individual grape. 



\subsection{Expectation Maximization and U-Net Reconstruction of CTIS images}\label{subsec:acq_nn}
The CTIS imaging system is described by the linear imaging equation that maps a datacube $\bm{f}$ to a CTIS image $\bm{g}$ using the system matrix $\bm{H}$~\cite{descour_computed-tomography_1995}: 
\begin{equation}
    \bm{g}=\bm{H} \bm{f} + \bm{n}
\end{equation}\label{eq:ctis}
where $\bm{g}$ is the vectorized pixels of the CTIS image, $\bm{f}$ is the vectorized voxels of the datacube, $\bm{n}$ is a random noise vector, and $\bm{H}$ is a 2D matrix. For our system, the $2048\times\SI{2048}{pixels}$ CTIS image is cropped to yield a vectorized $1910\times\SI{1910}{pixels}$ $\bm{g}$, while $\bm{f}$ is a vectorized $312\times312\times\SI{236}{voxels}$ datacube, resulting in a $(1910 \cdot 1910) \times (312 \cdot 312 \cdot 236)$ system matrix $\bm{H}$. The system matrix $\bm{H}$ incorporates the diffraction sensitivity (lens transmission, sensor response, and diffraction efficiency of the DOE), illumination, and vignetting.
For non-trivial system dimensions, the CTIS system is an underdetermined linear equation with no exact solutions since the inverse of $\bm{H}$ is nonexistent. Therefore, either iterative reconstruction algorithms such as the EM algorithm~\cite{descour_computed-tomography_1995, white_accelerating_2020} or CNNs~\cite{huang_application_2022,ahlebaek_hybrid_2023} are utilized to reconstruct datacubes by leveraging the high structural order from the diffraction grating in the CTIS images. The EM algorithm consists of an expectation step, where we compute $\hat{\bm{g}} = \bm{H} \hat{\bm{f}}^{(k)}$, and a maximization step, where we update $\hat{\bm{f}}^{(k)}$:
\begin{equation}
	\hat{\bm{f}}^{(k+1)}	 = \frac{\hat{\bm{f}}^{(k)}}{\sum_{i=1}^{q^2} H_{ij}}\odot \left( \bm{H}^T \frac{\bm{g}}{\bm{H}\hat{\bm{f}}^{(k)}}\right)
\end{equation}\label{eq:EM}
Here, $k$ is the iteration index, $\hat{\bm{f}}^{(k)}$ is the current estimate, $\sum_{i=1}^{q^2} H_{ij}$ is the sum of rows in $\bm{H}$, $\bm{H}^T$ is the transpose of $\bm{H}$, and $\odot$ denotes element-wise multiplication. This equation combines the expectation and maximization steps. We initialize with $\hat{\bm{f}}^{(0)} = \bm{H}^T\bm{g}$ and perform 20 iterations, as typically 10-30 are needed~\cite{white_accelerating_2020}. Both the $\bm{H}$ construction and $\bm{f}$ reconstruction are implemented in \texttt{MATLAB} using sparse matrix manipulations.

CNNs have previously been used for classification directly on CTIS images~\cite{douarre_ctis-net_2021} and to reconstruct datacubes from CTIS images~\cite{huang_application_2022,ahlebaek_hybrid_2023,mel2022joint,Wu,Zimmermann}. Here, we utilize a U-Net neural network to perform CTIS reconstructions. The architecture of the U-Net is equivalent to the U-Net presented in~\cite{huang_application_2022}, where the filter dimensions have been modified to accommodate the input $1910\times\SI{1910}{pixels}$ CTIS images and the output $312\times312\times \SI{45}{voxels}$ datacube. 

\begin{figure}
	\centering
	\begin{subfigure}[t]{\textwidth}
		\centering
		\includegraphics[width=1\textwidth]{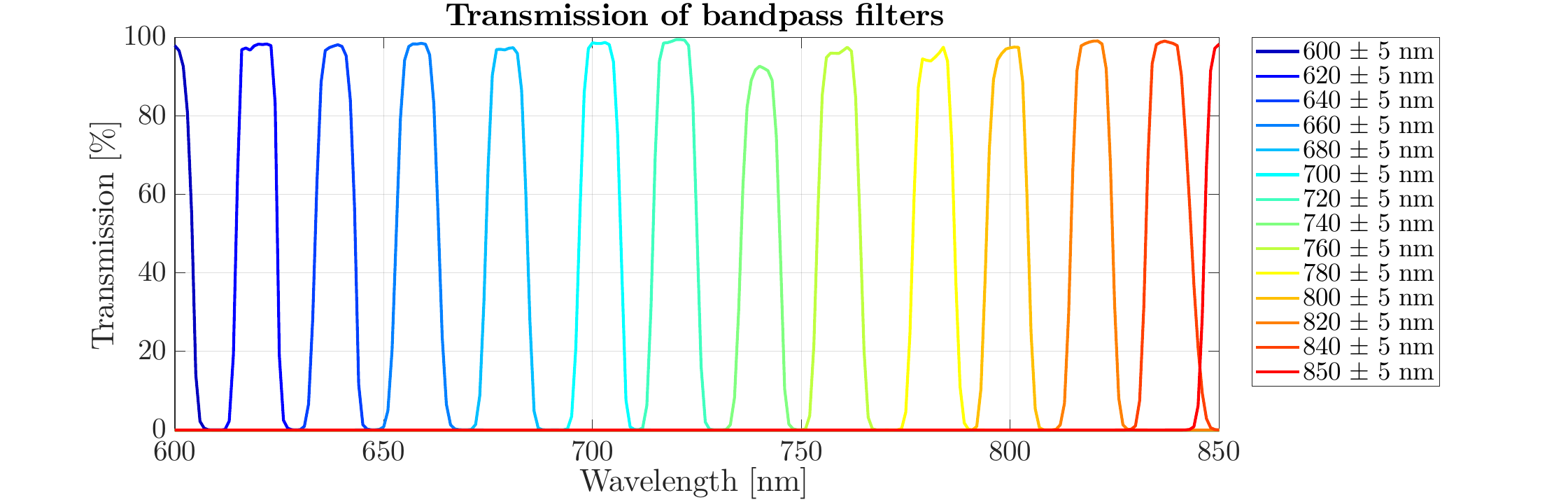}
		\caption{}
		\label{fig:filter_response}
	\end{subfigure}
	\hfill
	\begin{subfigure}[t]{0.582\textwidth}
		\centering
		\includegraphics[width=\textwidth]{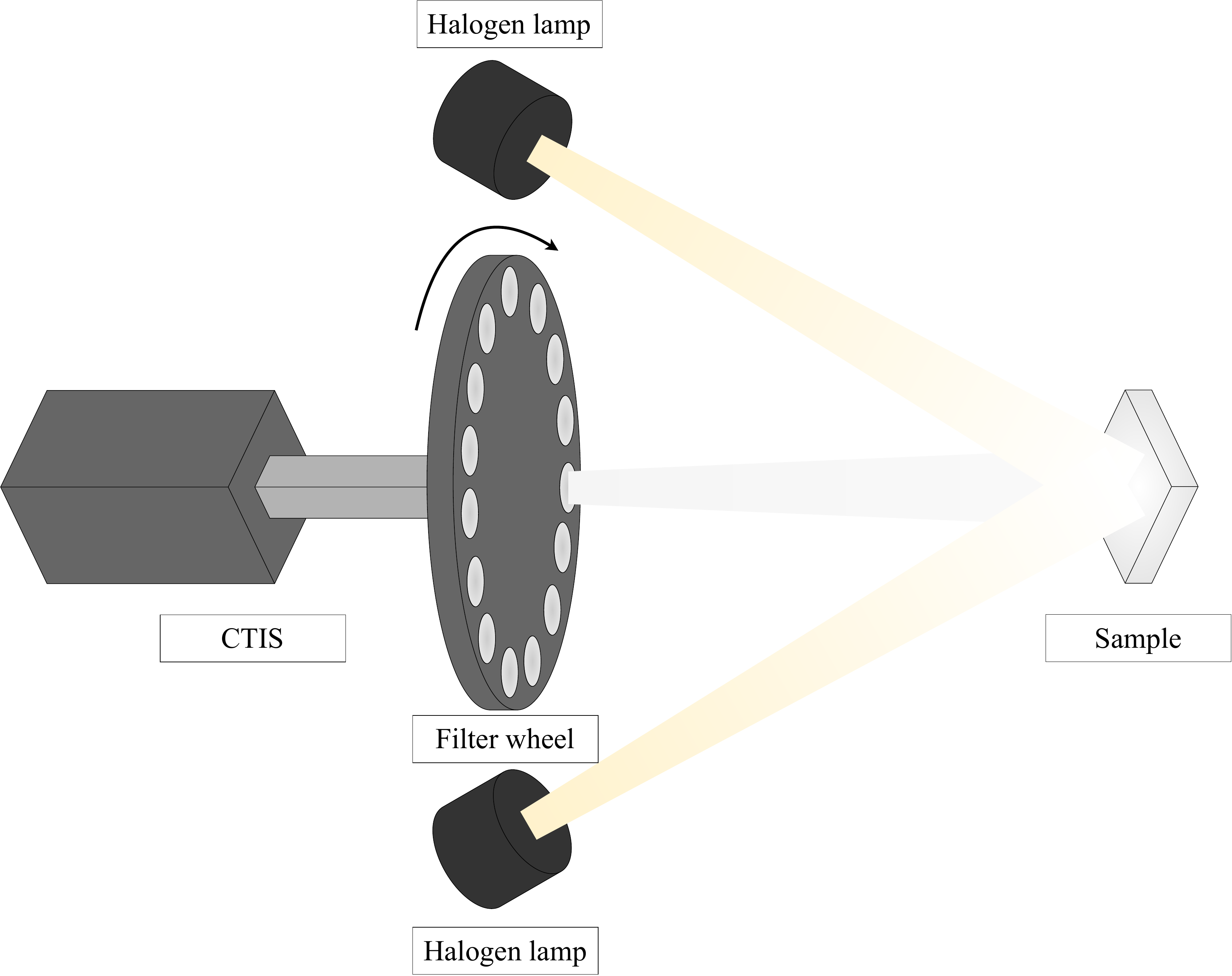}
		\caption{}
		\label{fig:filter_wheel_setup}
	\end{subfigure}
	\hfill
	\begin{subfigure}[t]{0.41\textwidth}
		\centering
		\includegraphics[width=\textwidth]{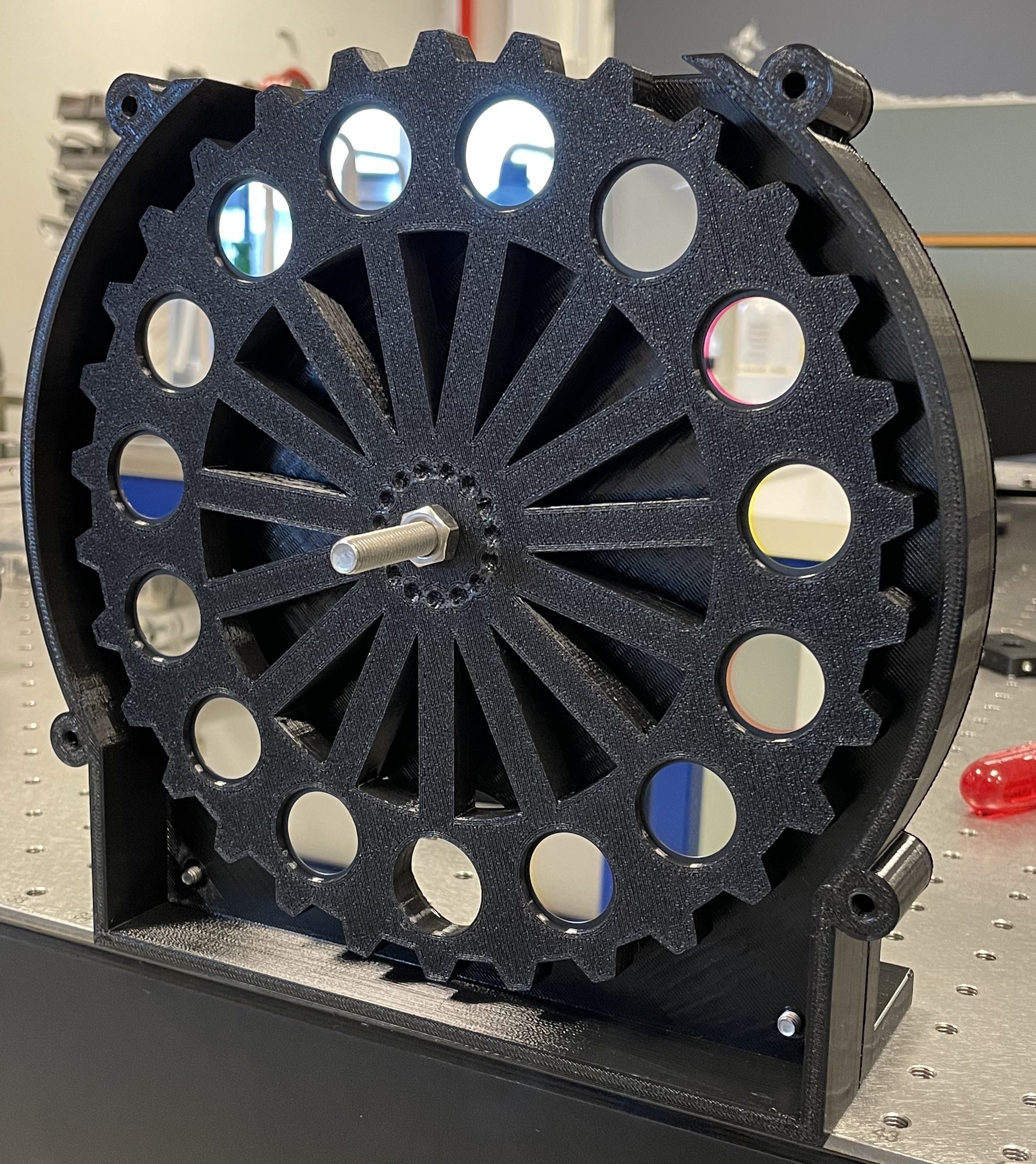}
		\caption{}
		\label{fig:filter_wheel}
	\end{subfigure}
	\caption{(a) Transmission of optical bandpass filters from 600-850 nm used for the acquisition of training data for the U-Net neural network. (b) Experimental setup for data acquisition of CTIS and multispectral images for ANN training. (c) Filter placement inside the filter wheel without the front cover.}
	\label{fig:filter_response_setup}
\end{figure}

To train the U-Net reconstruction network, we use an experimental setup consisting of the CTIS camera with a filter wheel positioned in front and two halogen lamps placed on either side of the camera at \SI{45}{\celsius} relative to the imaged object at a distance of \SI{60}{cm}. The DOE enclosure of the CTIS camera is replaced with a different DOE enclosure, allowing the DOE to be removed from the optical axis, resulting in a CTIS camera that only captures the central zeroth order due to the absence of diffraction. The filter wheel has 15 slots, containing 14 optical narrowband bandpass filters ranging from 600-850 nm with full-width-half-maximum bandwidths of \SI{10}{nm} (600 nm, 620 nm, 640 nm,..., 820 nm, 840 nm, 850 nm) and one empty slot for CTIS imaging. The transmission of the bandpass filters is illustrated in Fig.~\ref{fig:filter_response}. The bandwidth of the \SI{600}{nm} and \SI{850}{nm} bandpass filters are effectively halved by the \SI{600}{nm} and \SI{850}{nm} short- and longpass filters at the front of the CTIS camera, ensuring no overlap between the zeroth and first orders and that the first orders do not extend beyond the active area of the image sensor.

To capture training data, we first capture a CTIS image of the scene and a dark reference image. Then, the DOE is removed from the optical axis, and 14 bandpass filter images (with the same exposure time) are captured by manually rotating the filter wheel. To obtain a ground truth reflectance datacube, the FOV of the camera is filled with a RESTAN white reference target, and an additional 14 bandpass filter images are captured, which act as the white reference multispectral datacube. Dark reference images are acquired using the same camera settings, and the multispectral datacubes are transformed to reflectance by dividing the dark frame-subtracted raw bandpass filter images by the dark frame-subtracted white reference images. 

A deep neural network (DNN) is utilized to interpolate 14-channel datacubes into 45-channel datacubes. The DNN is trained using datacubes acquired from a HSI line scan system with 180 spectral channels in the wavelength range of 600–850 nm. These datacubes are processed by convolving their spectra with bandpass filter responses (Fig.~\ref{fig:filter_response}), resulting in 14-channel spectra. The DNN is then tasked with reconstructing the original 180-channel spectra from the reduced 14-channel spectra. To enhance spectral variability, pseudo-spectra generated from Gaussian mixtures are included. After reconstructing the 180-channel spectra, the number of spectral channels is reduced to 45 using bicubic interpolation. Once trained, the DNN is applied to reconstruct 45 spectral channels from the captured multispectral datacubes with 14 channels, producing image pairs of CTIS images and their corresponding 45-channel datacubes.

The U-Net network used for CTIS reconstruction is trained on reflectance datacubes and corresponding dark-reference-subtracted CTIS images. In total, 389 pairs of training data are utilized to initially train the U-Net. To further enhance the training set, 600 additional line scan datacubes (downsampled to 45 spectral channels) are incorporated. From these, 24,010 pseudo-datacubes and their corresponding pseudo-CTIS images are generated with an updated system matrix, following the approach outlined in~\cite{huang_application_2022}. This expanded dataset is split into training, validation, and test sets in proportions of \SI{70}{\percent}, \SI{15}{\percent}, and \SI{15}{\percent}, respectively, resulting in 17,135, 3,631, and 3,631 image pairs for training, validation, and testing.

The U-Net used for CTIS reconstruction is trained on reflectance datacubes and dark-reference-subtracted CTIS images. Initially, 389 pairs of training data are used to train the U-Net. To further augment the training set, 600 additional line scan datacubes (downsampled to 45 spectral channels) are added. From these, 24,010 pseudo-datacubes and their corresponding pseudo-CTIS images are generated using an updated system matrix, following the approach in Huang et al.~\cite{huang_application_2022}. This expanded dataset is divided into training, validation, and test sets, with proportions of \SI{70}{\percent}, \SI{15}{\percent}, and \SI{15}{\percent}, respectively, resulting in 17,135, 3,631, and 3,631 image pairs for training, validation, and testing.



\section{Results}\label{Results}

\begin{figure}
    \centering
    \begin{subfigure}[b]{\textwidth}
        \centering
        \includegraphics[width=1\textwidth]{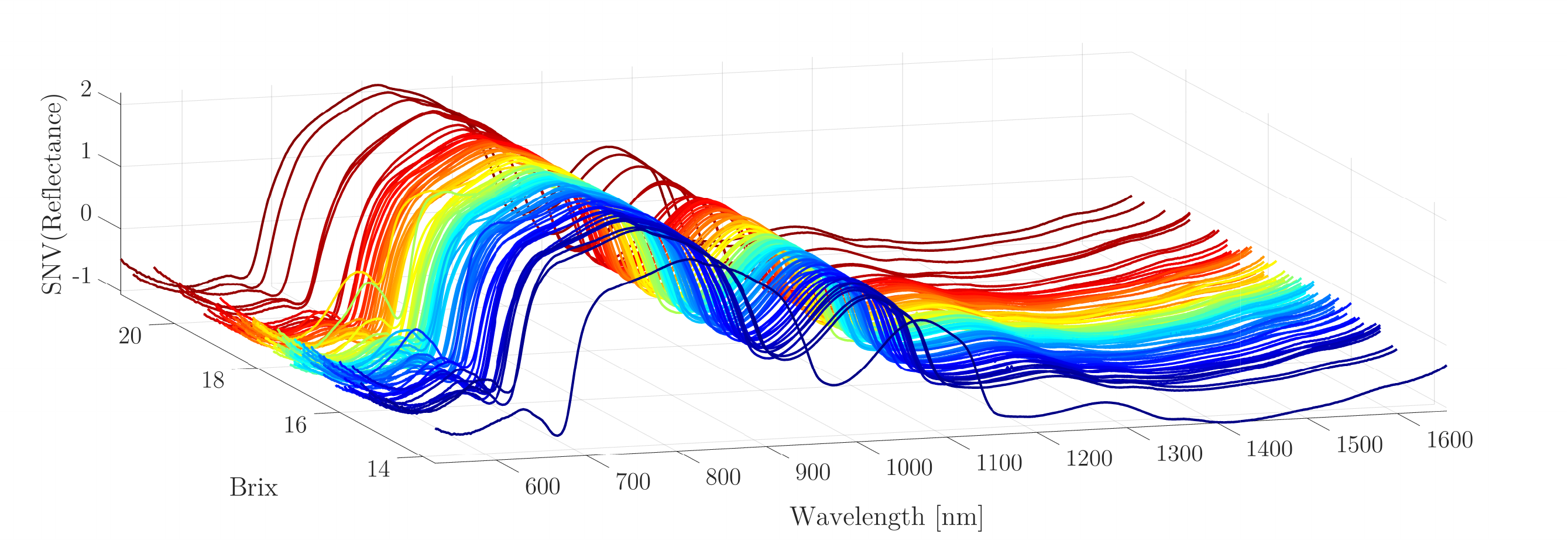}
        \caption{Line scan (Buteo) 532-1655 nm}
        \label{fig:buteo_mean_spectra_buteo}
    \end{subfigure}
    \hfill
    \begin{subfigure}[b]{0.49\textwidth}
        \centering
        \includegraphics[width=1\textwidth]{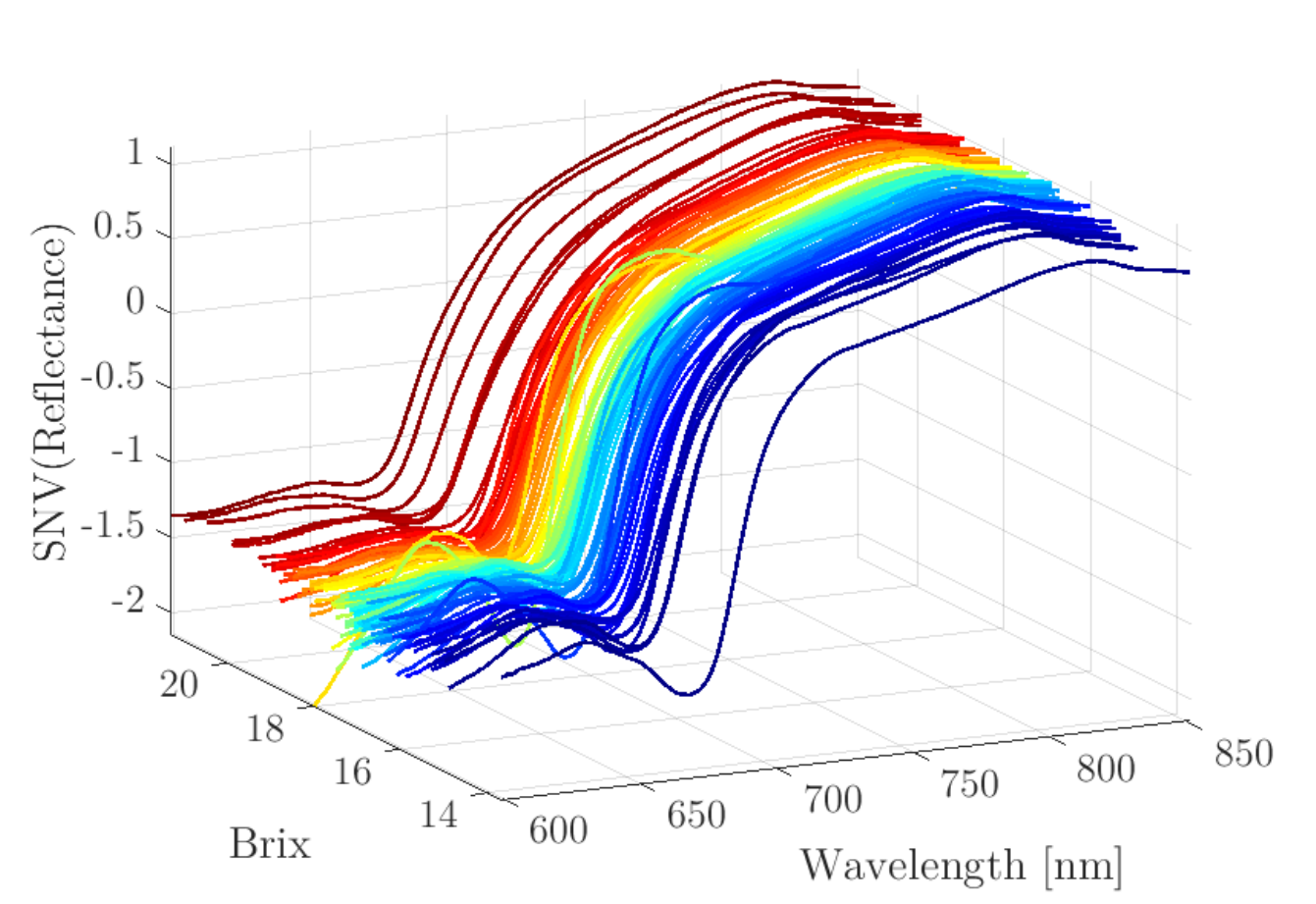}
        \caption{Line scan (Buteo)  600-850 nm}
        \label{fig:ctis_mean_spectra_ctis}
    \end{subfigure}
    \hfill
    \begin{subfigure}[b]{0.49\textwidth}
        \centering
        \includegraphics[width=1\textwidth]{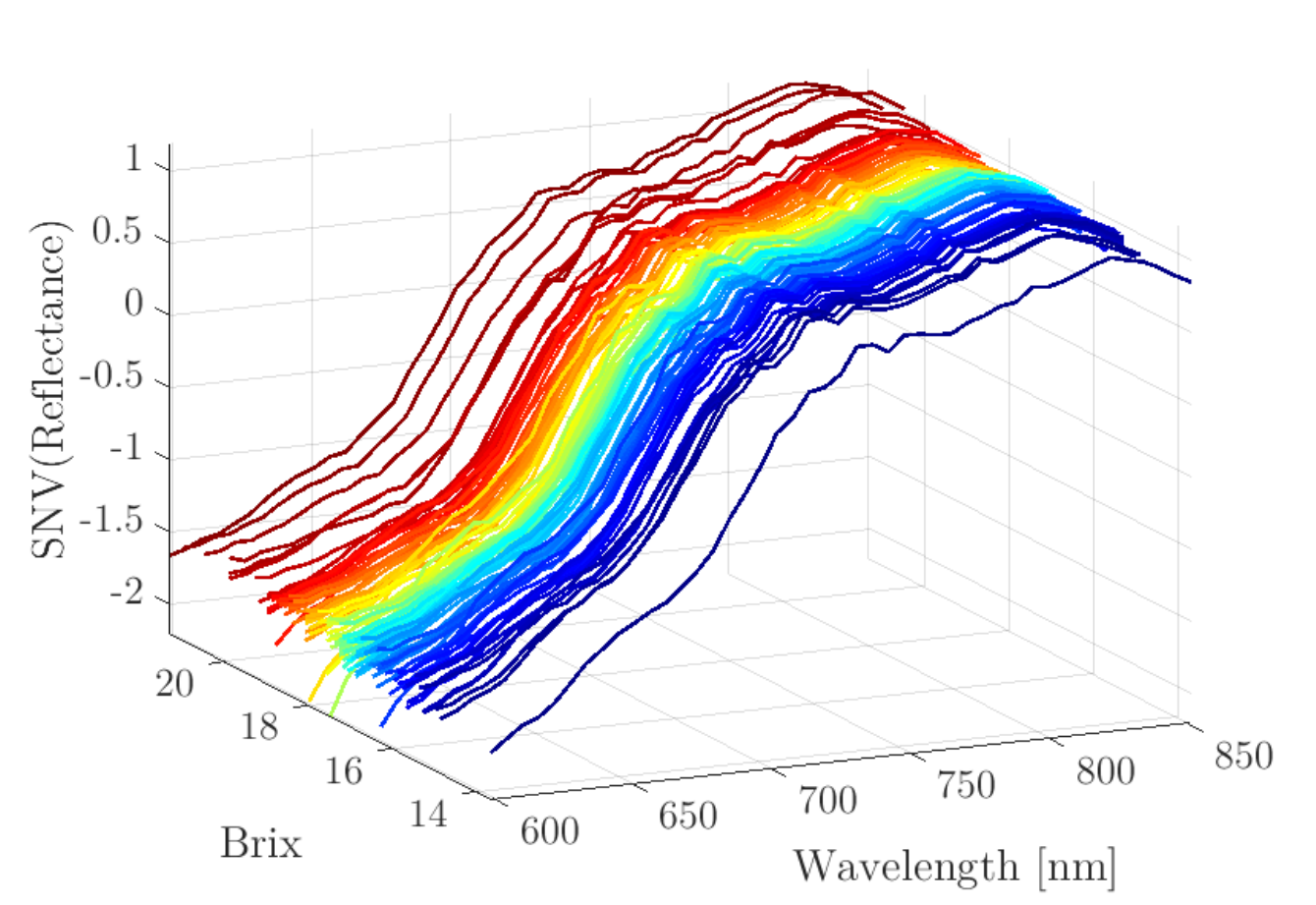}
        \caption{Snapshot (CTIS) 600-850 nm}
        \label{fig:ctis_mean_spectra_600_850nm}
    \end{subfigure}
    \caption{Standardized mean spectra (expressed by SNV) of grapes calculated from the acquired datacubes. (a) Line scan datacube from 523-\SI{1655}{nm}, (b) line scan datacube from 600-\SI{850}{nm}, and (c) snapshot datacube from 600-\SI{850}{nm} (U-Net reconstructions).}
    \label{fig:Mean_spectra}
\end{figure}

To compare the spectra of the \textit{Sheegene 20} grapes, a Standard normal variate (SNV) transformation is performed on their respective spectral information.
This will be referred to as the standardized mean reflectance spectra. This was performed for both the line scan and snapshot hyperspectral cameras which is visualized in Fig. \ref{fig:Mean_spectra}\textcolor{blue}{a-c}, where the colors indicate the measured \textdegree Brix value. The full spectral range of 445-1710 nm of the hyperspectral line scan system is reduced to 532-1655 nm by removing the ends of the spectrum due to low signal-to-noise ratio. Thus, Fig. \ref{fig:buteo_mean_spectra_buteo} shows the spectral range from 532-1655 nm of the line scan datacube, while Fig. \ref{fig:ctis_mean_spectra_ctis} and Fig. \ref{fig:ctis_mean_spectra_600_850nm} show the reduced spectral range of 600-850 nm for the line scan and snapshot datacubes, respectively. 

For each of the three cases (line scan range from 532-1655 nm, reduced line scan from 600-850 nm, and CTIS from 600-850 nm), a predictive PLSR model that correlates the spectral information with the \textdegree Brix and pH values is constructed. 

All models are trained using 10 K-fold cross-validation (CV), and the number of PLS components is chosen to minimize the predictive error of the models. For each model, the cross-validated $R^2$, $RMSECV$ and $MAECV$ are determined:
\begin{equation}
    RMSECV = \sqrt{\frac{1}{n} \sum_{i=1}^n\left(y_i - \hat{y_i} \right)^2 } \qquad \qquad MAECV = \frac{1}{n} \sum_{i=1}^n\left|y_i - \hat{y_i}\right|
\end{equation}
where $y_i$ and $\hat{y_i}$ are the measured and predicted response variables, respectively. The performance of each model is summarized in Table \ref{tab:results}.

\begin{table}[h]
\rowcolors{2}{gray!25}{white}
    \centering
    \resizebox{\linewidth}{!}{%
\begin{tabular}{l|lcccccccc}
\textbf{System} &
  \textbf{Spectral channels} &
  \multicolumn{1}{l}{\textbf{Spectral range}} &
  \multicolumn{1}{l}{\textbf{Comp.}} &
  \multicolumn{1}{l}{\textbf{$\mathbf{R^2_{Brix}}$}} &
  \multicolumn{1}{l}{\textbf{$\mathbf{R^2_{pH}}$}} &
  \multicolumn{1}{l}{\textbf{$\mathbf{MAECV_{Brix}}$}} &
  \multicolumn{1}{l}{\textbf{$\mathbf{MAECV_{pH}}$}} &
  \multicolumn{1}{l}{\textbf{$\mathbf{RMSECV_{Brix}}$}} &
  \multicolumn{1}{l}{\textbf{$\mathbf{RMSECV_{pH}}$}} \\ \hline
Buteo &  800 & 532-1655 nm & 14 & 0.80          & 0.52          & 0.53          & 0.054          & 0.73          & 0.072          \\
Buteo &  180 & 600-850 nm  & 7 & 0.56           & 0.27          & 0.81 & 0.056 & 1.02 & 0.074 \\
CTIS & 236 & 600-850 nm & 7 & 0.51 & 0.34 & 0.83 & 0.053 & 1.08 &  0.071  \\
CTIS  &  45  & 600-850 nm  & 7 & 0.52           & 0.29          & 0.79 & 0.057 & 1.03 & 0.070
\end{tabular}}
    \caption{Overview of the performance of the PLSR models for line scan (523-1655 nm and 600-850 nm) and CTIS (600-850 nm) for the \textit{Sheegene 20 grapes}. 10 K-fold cross-validation is used for all models.}
    \label{tab:results}
\end{table}

\begin{figure}
    \centering
    \includegraphics[width=1\textwidth]{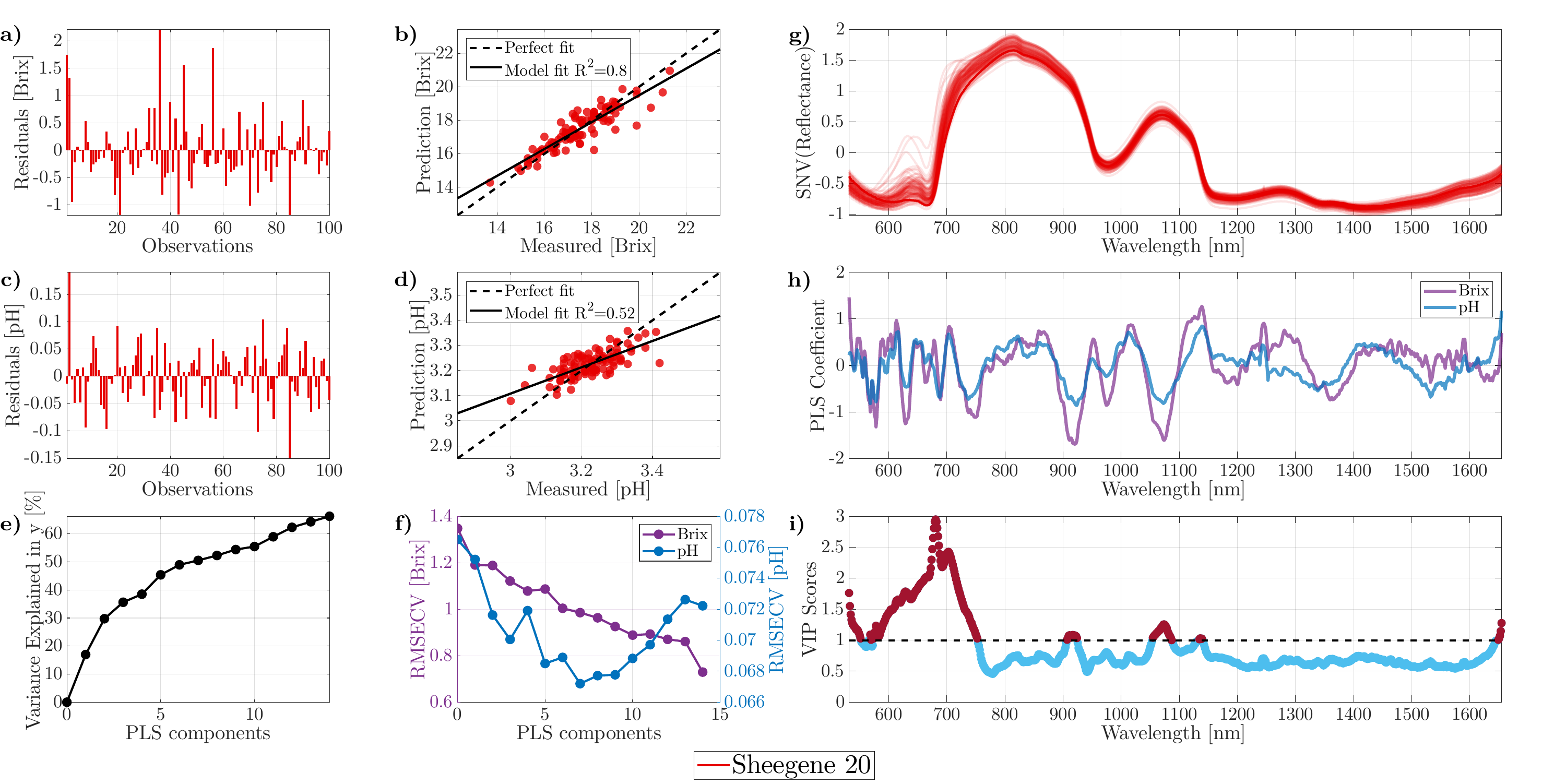}
    \caption{PLSR results for VisSWIR line scan images in the wavelength range from 532-1655 nm. 10 K-fold cross-validation was performed with 14 PLS components, resulting in a $\text{MAECV}_\text{Brix}=0.53 \text{ Brix}$ and $\text{MAECV}_\text{pH}=0.054 \text{ pH}$.}
    \label{fig:buteo_515-1640nm_1}
\end{figure}

Fig. \ref{fig:buteo_515-1640nm_1} presents the results for the PLSR model applied to the line scan datacube with a spectral range of 532-1655 nm. The standardized mean spectra of each grape, averaged over the three rotations, are shown in Fig.~\ref{fig:buteo_515-1640nm_1}\textcolor{blue}{g}.
The spectra display expected reflectance peaks near 650 nm, attributed to red pigments such as anthocyanins and flavonoids in \textit{Sheegene 20}, and around 800-850 nm, related to the grape sugar content~\cite{baiano_application_2012,s23031065}. 
Additionally, notable absorption features are observed near the wavelengths 680 nm, 970 nm, 1200 nm,
and 1440 nm, consistent with findings from~\cite{s23031065}. 
Spectral absorption features at 980 nm (due to sugars and water) and 1150 nm (due to anthocyanins) are also anticipated~\cite{DOSSANTOSCOSTA2019166}.

The residuals and the PLSR model fit for the two response variables (Fig. \ref{fig:buteo_515-1640nm_1}\textcolor{blue}{a-d}) demonstrate strong predictive performance with $R_{Brix}^2=0.80$ and $R_{pH}^2=0.52$ comparable to the results reported in e.g.~\cite{baiano_application_2012}. 
The model captures most of the variance in the response variables (Fig. \ref{fig:buteo_515-1640nm_1}\textcolor{blue}{e}), and the cross-validation root-mean-squared-error (RMSE) for \textdegree Brix and pH (Fig. \ref{fig:buteo_515-1640nm_1}\textcolor{blue}{f}) decreases up to 14 PLS components before increasing (not shown), indicating potential overfitting with additional components. The corresponding $RMSECV$ and $MAECV$ for \textdegree Brix and pH are $RMSECV_{Brix} = 0.73, RMSECV_{pH} = 0.072$ and $MAECV_{Brix} = 0.53, MAECV_{pH} = 0.054$.

The PLSR coefficients for \textdegree Brix and pH, and the variable importance projection (VIP) scores, highlight the spectral regions that most influence the PLSR models.  VIP scores greater than 1 indicate significant spectral regions around 532-775 nm and 900-1150 nm

\begin{figure}[htp]
    \centering
    \includegraphics[width=1\textwidth]{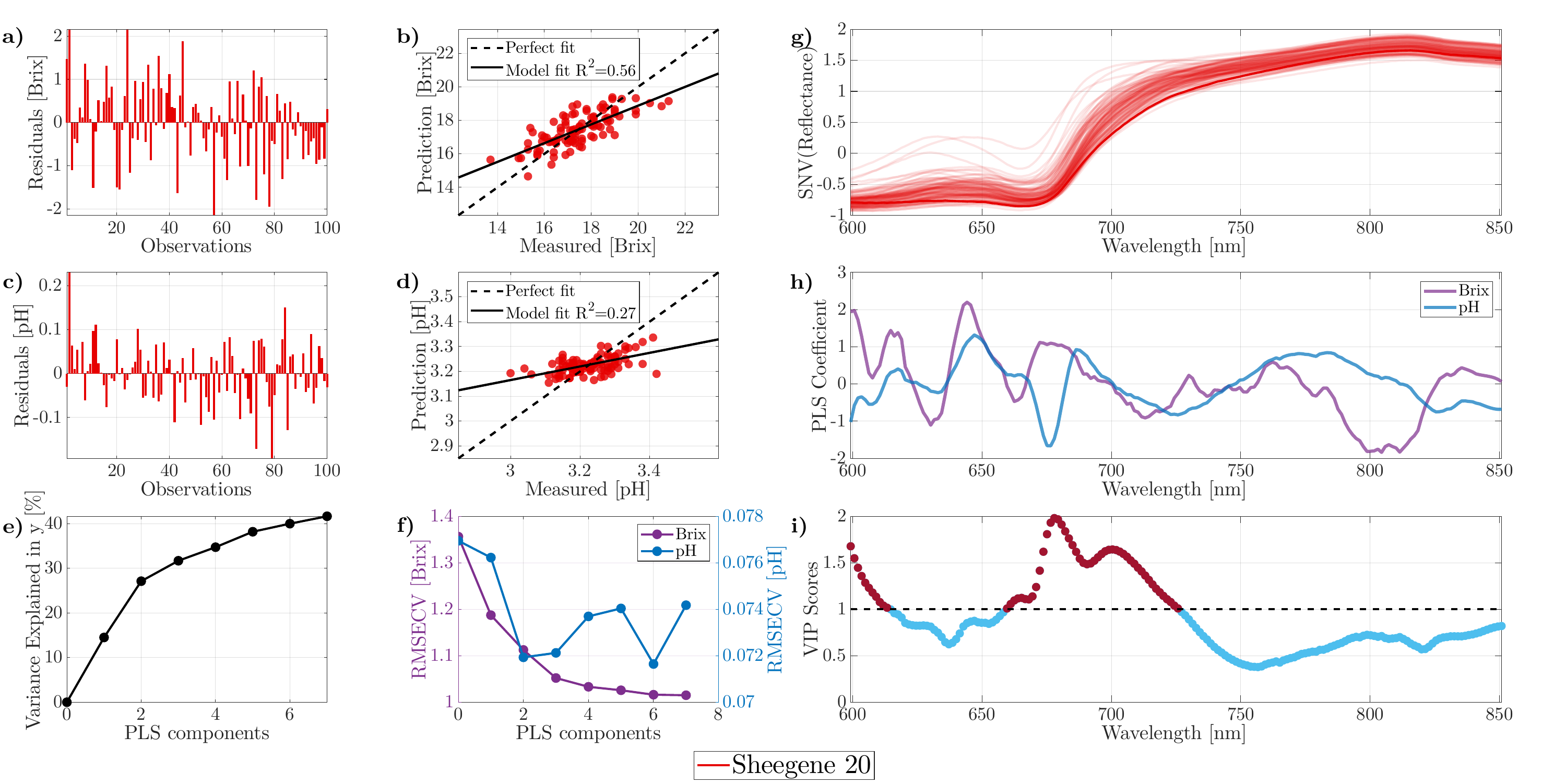}
    \caption{PLSR results for VisSWIR line scan images reduced to the CTIS wavelength range from 600-850 nm. 10 K-fold cross-validation was performed with 7 PLS components, resulting in a $\text{MAECV}_\text{Brix}=0.81\text{ Brix}$ and $\text{MAECV}_\text{pH}=0.056 \text{ pH}$. }
    \label{fig:buteo_600-850nm_1}
\end{figure}

To assess the performance of the line scan and snapshot cameras, the line scan datacubes were restricted to the 600–850 nm spectral range, and new PLSR models were developed (Fig. \ref{fig:buteo_600-850nm_1}). As expected, limiting the spectral range led to a decline in model performance. The PLSR model for the reduced spectral range produced an RMSECV of 1.02 for Brix and 0.074 for pH, with MAECV values of 0.81 for Brix and 0.056 for pH. While the prediction error for pH increased slightly, the error for Brix showed a more significant rise due to the constrained spectral range.

\begin{figure}[htp]
    \centering
    \includegraphics[width=1\textwidth]{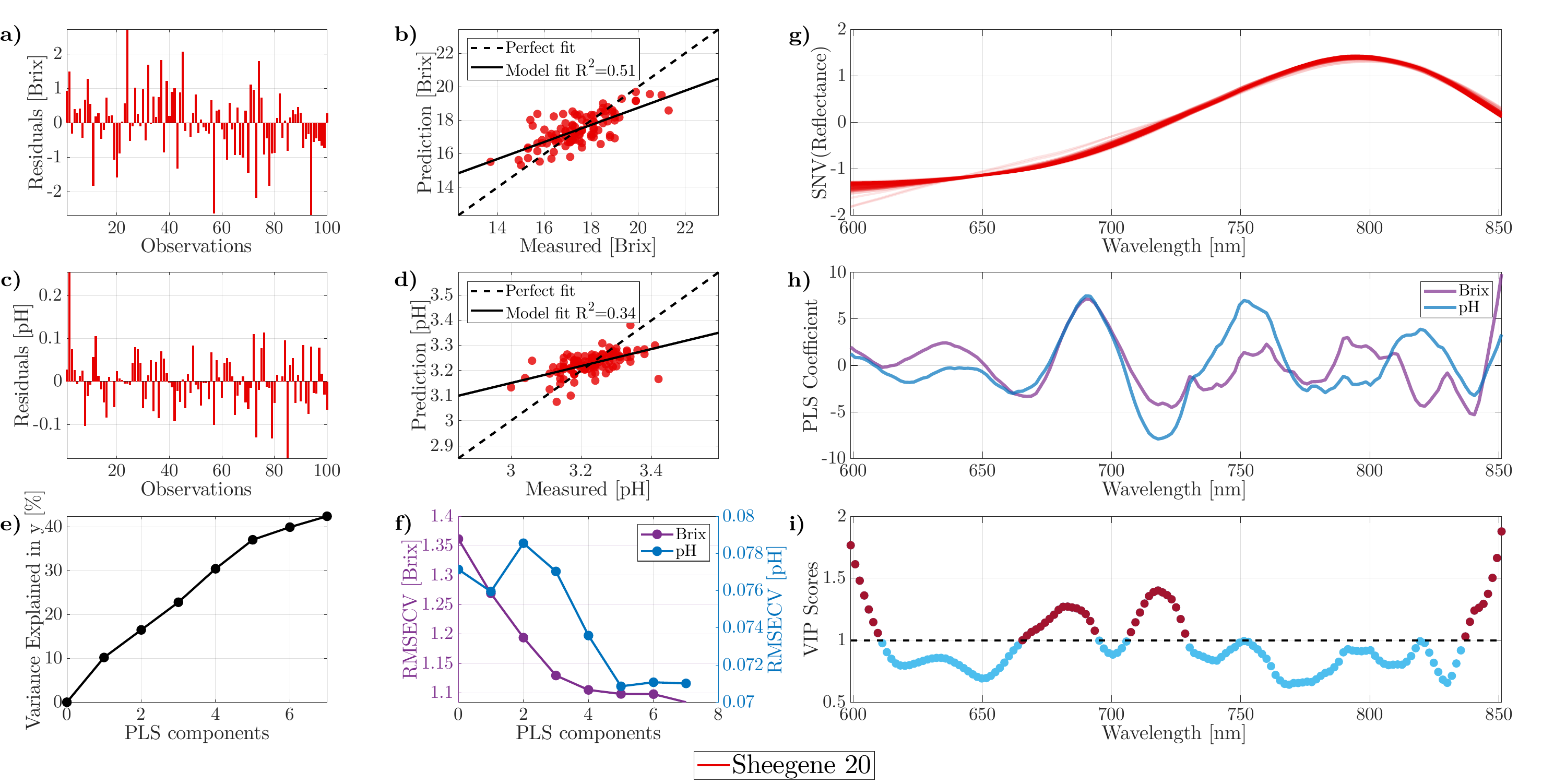}
    \caption{PLSR results for EM CTIS reconstructions in the wavelength range from 600-850 nm. 10 K-fold cross-validation was performed with 7 PLS components, resulting in a $\text{MAECV}_\text{Brix}=0.83\text{ Brix}$ and $\text{MAECV}_\text{pH}=0.053 \text{ pH}$.}
    \label{fig:ctis_600-850nm_4}
\end{figure}

The CTIS images were then evaluated. First, datacubes reconstructed using the EM algorithm (which operates without training) with 236 spectral channels were analyzed (Fig. \ref{fig:ctis_600-850nm_4}). The standardized reflectance spectra from these datacubes exhibited fewer features than the line scan spectra, appearing smoothed with less distinction between different grape varieties. The PLSR model based on these datacubes achieved an RMSECV of 1.08 for Brix and 0.071 for pH, with MAECV values of 0.83 for Brix and 0.053 for pH—comparable to the line scan model in the 600–850 nm range.

\begin{figure}[htp]
    \centering
    \includegraphics[width=1\textwidth]{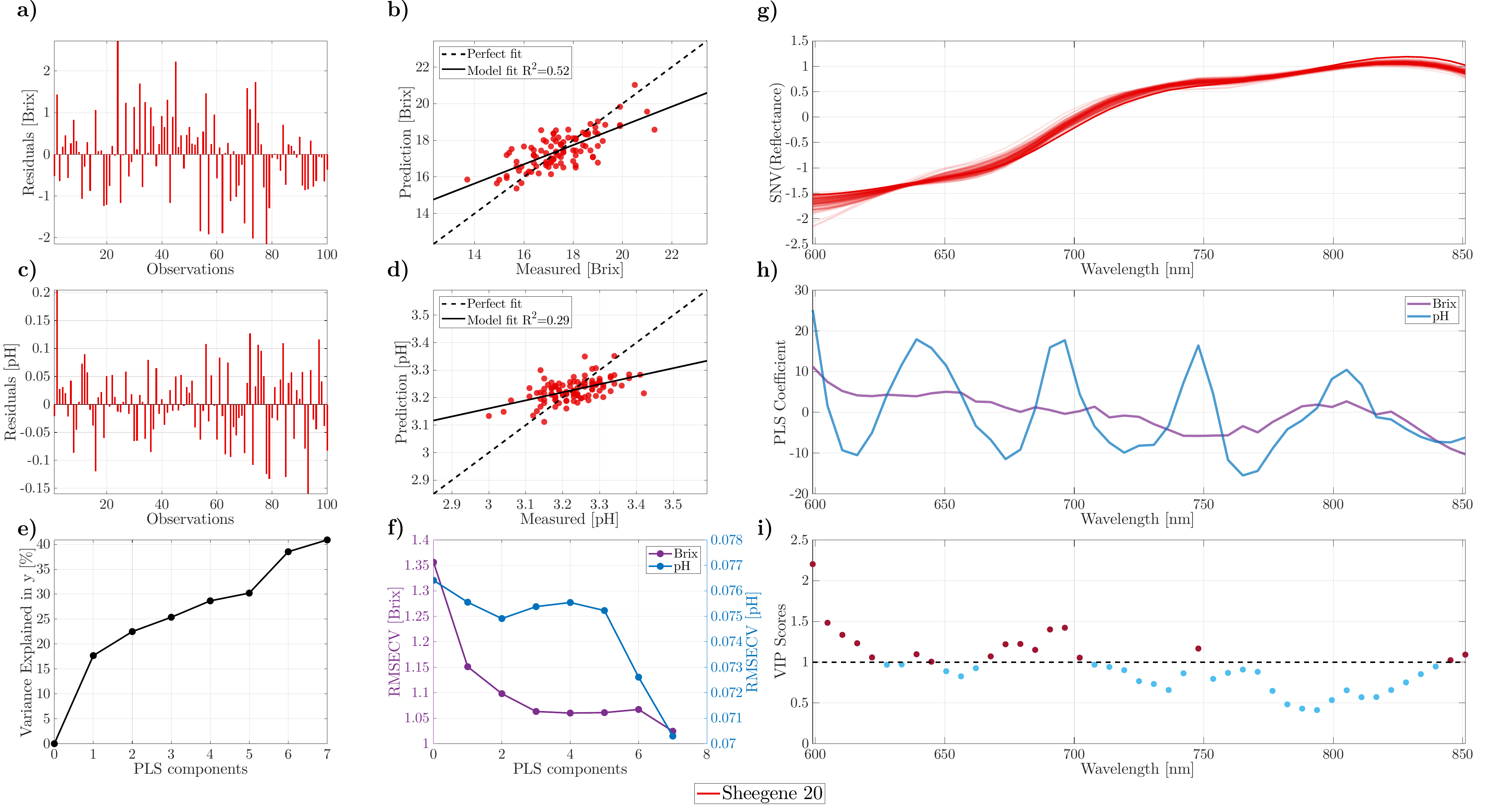}
    \caption{PLSR results for U-Net CTIS reconstructions in the wavelength range from 600-850 nm. 10 K-fold cross-validation was performed with 7 PLS components, resulting in a $\text{MAECV}_\text{Brix}=0.79\text{ Brix}$ and $\text{MAECV}_\text{pH}=0.057 \text{ pH}$.}
    \label{fig:ctis_600-850nm_3}
\end{figure}

Finally, a PLSR model was constructed from datacubes reconstructed by the U-Net with 45 spectral channels (Fig.~\ref{fig:ctis_600-850nm_3}). The reconstructed spectra, as seen in Fig.~\ref{fig:ctis_600-850nm_3}\textcolor{blue}{g}, captured more features and showed better alignment with the line scan system compared to the EM-reconstructed spectra. However, the spectra appeared more compressed between individual grapes and smeared across the spectral range, likely due to the U-Net reconstruction algorithm. Although the U-Net was generically trained and not yet optimized for grape analysis, these initial results are promising. The resulting RMSECV values were 1.03 for Brix and 0.070 for pH, with MAECV values of 0.79 for Brix and 0.057 for pH, closely matching those from the reduced line scan model.

\section{Discussion}\label{sec:discussion}
The results presented in Sec.~\ref{Results} depend on the choices made regarding image preprocessing, reconstruction (specifically for CTIS), and analysis. Building on the early study by Baiano et al.~\cite{baiano_application_2012}, a simple PLSR model with minimal image preprocessing was developed to determine the pH and \textdegree Brix of Sheegene 20 grapes from hyperspectral line scan and CTIS snapshot images. In the aforementioned study, it was found that more extensive preprocessing did not significantly affect the results, and a similar outcome is anticipated for the current analysis.

For the predictive method, a standard PLSR model was selected. While future refinements to this model could be beneficial, a comparison of 16 different linear and non-linear predictive methods for determining grape quality parameters, as reported by Gomes et al.~\cite{gomes_determination_2021}, indicated that PLSR was among the most effective. Therefore, only marginal improvements are expected from similar studies in the future.

The U-Net employed for CTIS image reconstruction was initially trained on a general dataset of CTIS-datacube pairs, consisting of 389 images that included scenes and objects with significant variations in geometry and spectral properties. It is anticipated that a dedicated network trained specifically for grape image reconstruction will yield substantial improvements in spectral quality. However, achieving a balance between the network's ability to reconstruct spectra from specific grape varieties and its generalizability to other varieties and field conditions will require careful and extensive research, which will be addressed in future studies.

The measured dataset, as illustrated in Fig. \ref{fig:buteo_515-1640nm_1}\textcolor{blue}{g}, retains outliers intentionally during model development. This approach increases the likelihood that the models will perform well not only under ideal laboratory conditions with store-bought table grapes but also in-field measurements, where greater grape variability exists.

In a study by Shirashi and Mikio~\cite{Shirashi_Mikio}, the variation in both acidity and SSC among four different cultivars over three successive years was investigated, although \textit{Sheegene 20} was not included in this analysis. This study reported yearly SSC variation ranging from 0.61 to 2.20 \textdegree Brix within the same cultivar. Some cultivars were noted to have low SSC values of 14.3 $\pm$ 0.81 \textdegree Brix, 14.3 $\pm$ 0.61 \textdegree Brix, and 14.7 $\pm$ 1.80 \textdegree Brix. Using these findings as a benchmark for evaluating the performance of the PLSR models presented here, the RMSECV for the non-restricted line scan system (532–1655 nm with 800 bands) aligns closely with the lower end of the observed variation. In contrast, the RMSECV for both the restricted line scan and CTIS systems, while slightly higher, still falls within the observed variation range.

The results of this study suggest several intriguing directions for future research. Firstly, training the neural network-based reconstruction specifically on grapes could enhance performance by incorporating a larger dataset and additional grape varieties, thereby improving generalizability and spectral sensitivity of the CTIS data. Other promising approaches include training a model directly on the 2D CTIS images, which would effectively bypass reconstruction and its associated errors, and testing the CTIS camera for field imaging applications.





\section{Conclusion}\label{sec:conclusion}

This study presents a comparative investigation of a newly developed snapshot hyperspectral camera based on Computed Tomography Imaging Spectroscopy (CTIS) and the state-of-the-art hyperspectral line scan system, \textit{Buteo}, from Newtec. The CTIS system captures 2D diffraction images that are reconstructed using both the iterative expectation maximization (EM) algorithm and an U-Net reconstruction network.

The CTIS camera acquires diffraction images measuring \(1910 \times 1910\) pixels, producing datacubes of dimensions \(312 \times 312 \times 236\) for EM reconstructions and \(312 \times 312 \times 45\) for U-Net reconstructions, covering the spectral range of 600-850 nm. In contrast, the line scan system captures images of dimensions \(1296 \times m \times 900\), where \(m\) represents the scan length in pixels, across a broader spectral range of 445-1710 nm. However, due to low signal-to-noise ratios, the effective "full" range for the line scan system is reduced to 532-1655 nm.

Despite its narrower spectral range, the relevance of the CTIS system lies in its lower cost, capability to capture snapshot hyperspectral images (which reduces motion-related errors), and its portability for field applications. To ensure a fair comparison between the two systems, both were constrained to the spectral range of 600-850 nm. A PLSR model was constructed from the intermediate data, as detailed in the results section (Sec. \ref{Results}), to assess the determination of Soluble Solids Content (SSC) and acidity in 100 \textit{Sheegene 20} table grapes, measured in °Brix and pH, respectively.

The PLSR model based on the full spectral range of the line scan system demonstrated strong predictive performance with \(RMSECV_{Brix} = 0.73\), \(RMSECV_{pH} = 0.072\), \(MAECV_{Brix} = 0.53\), and \(MAECV_{pH} = 0.054\). The \(R_{Brix}^2\) and \(R_{pH}^2\) values were 0.80 and 0.52, respectively, indicating robust spectral information for SSC, while the evidence for pH prediction was less convincing. Nevertheless, the performance for SSC aligns well with previous findings in the literature, such as those reported by Baiano et al. and others~\cite{baiano_application_2012,s23031065,DOSSANTOSCOSTA2019166}.

When the spectral range for the line scan system is reduced, performance diminishes to \(RMSECV_{Brix} = 1.02\) and \(RMSECV_{pH} = 0.074\), with corresponding \(R^2\) values of 0.56 and 0.27, and \(MAECV_{Brix} = 0.81\) and \(MAECV_{pH} = 0.056\). This illustrates a notable decline in predictive accuracy, yet SSC predictions remain relatively robust.

For the CTIS images, the PLSR model based on EM reconstructions yielded \(RMSECV_{Brix} = 1.08\) and \(RMSECV_{pH} = 0.071\), with \(R^2\) values of 0.51 and 0.34, and \(MAECV_{Brix} = 0.83\) and \(MAECV_{pH} = 0.053\). The U-Net reconstruction resulted in \(RMSECV_{Brix} = 1.03\) and \(RMSECV_{pH} = 0.070\), with \(R^2\) values of 0.52 and 0.29, respectively, alongside \(MAECV_{Brix} = 0.79\) and \(MAECV_{pH} = 0.057\). Similar to the reduced line scan system, SSC predictions yielded satisfactory results, though the \(R^2\) values for pH were low.

In summary, while the reconstructed datacubes from the CTIS snapshot system do not fully recover all spectral features, they perform comparably to the line scan system with reduced spectral range for SSC and acidity predictions. Given the CTIS camera's portability and cost advantages, it demonstrates significant potential for in-field applications under spectrally comparable conditions.

Future work should focus on improving the training of the image reconstruction algorithm for the CTIS camera, which may enhance the results obtained in this study. Overall, this research marks a promising first step toward optimizing the CTIS camera for practical applications in out-of-the-lab-application.

\section*{Acknowledgements}
MSP acknowledges partial funding from the Innovation Fund Denmark (IFD) under File No. 1044-00053B. We acknowledge partial funding from Food $\&$ Bio Cluster Denmark and the SDU Climate Cluster.

 \bibliographystyle{elsarticle-num} 
 \bibliography{ref}





\end{document}